\documentclass[event]{sigirforum}

\def\pubissue{Vol. 58 No. 1 -- June 2024}

\def\eventdate{March 26, 2026.}
\def\eventurl{https://proactive-chiir.github.io/}

\begin{document}
\title{Report on The 1st Workshop on Human-Centered Proactive and Personalized Agents for Interactive Information Access at CHIIR 2026}

\authors{
\author[kaur13@cs.washington.edu]{Kirandeep Kaur}{University of Washington}{Seattle\, USA}
\and
\author[guptavinayak51@gmail.com]{Vinayak Gupta}{University of Washington}{Seattle, USA}
\and
\author[tanya.roosta@gmail.com]{Tanya Roosta}{AMD \& UC Berkeley}{USA}
\and
\author[soymadhu@gmail.com]{Madhura Raju}{TikTok}{San Jose, USA}
\and \hfill
\author[grace.yang@georgetown.edu]{Grace Hui Yang}{Georgetown University}{Washington, D.C, USA}
\and
\author[chirags@uw.edu]{Chirag Shah}{University of Washington}{Seattle, USA}}

\maketitle 
\begin{abstract}
Interactive information access is increasingly moving beyond reactive query-response paradigms toward agentic systems that can personalize interaction, retain context, infer latent needs, recommend next steps, and initiate support. This shift creates new opportunities for adaptive and context-aware assistance, while also raising important questions about autonomy, privacy, trust, transparency, user welfare, and evaluation. The First Workshop on Human-Centered Proactive and Personalized Agents for Interactive Information Access provided an interdisciplinary forum for examining these questions across information retrieval, human-computer interaction, dialogue systems, AI ethics, cognitive science, learning technologies, and human-centered AI. Through invited talks, paper presentations, and open discussion, the workshop engaged with topics including calibrated initiative, knowledge-gap navigation, long-term memory, value-sensitive design, implicit personalization, AI-mediated care, proactive dialogue, and evaluation beyond task accuracy. A central theme across the workshop was that proactivity should not be understood only as earlier action or improved prediction, but as a form of initiative that must be appropriately timed, transparent, contestable, and aligned with user goals. This report summarizes the workshop and synthesizes the research challenges it surfaced for designing proactive and personalized agents in interactive information access.
\end{abstract}

\section{Introduction}

A persistent challenge in information access is that users do not always know how to formulate what they need~\cite{kaur2026knowing}. Classical information retrieval has long recognized this problem: users may begin from anomalous states of knowledge, revise their goals during interaction, and move through nonlinear patterns of searching, browsing, and sensemaking \cite{belkin1980ask,bates1989berrypicking}. Conversational information seeking extends this view by treating information access as an interactive process involving clarification, refinement, recommendation, and dialogue rather than a single query-response exchange \cite{zamani2023conversational}. The recent emergence of agentic AI systems makes this long-standing insight newly consequential. If users often cannot fully specify their needs in advance, then future information access systems may need to do more than wait for explicit queries.

Proactive and personalized agents represent one response to this challenge. Such systems can retain context across sessions, infer preferences, identify possible gaps in a user's framing, recommend next steps, and initiate support before a request is fully articulated. This shift creates opportunities for more adaptive forms of search, recommendation, learning support, and decision-making~\cite{10.1145/3774778}. It also changes the design problem. The central question is no longer only whether a system can retrieve relevant information or generate a useful response, but whether it can decide when initiative is warranted, what kind of intervention is appropriate, and how that intervention should remain accountable to the user.

This is a difficult balance~\cite{kaur2026proper}. A proactive intervention may help a user notice a missing constraint, avoid premature commitment, discover an adjacent concept, or make progress toward a long-term goal. The same intervention, however, may become intrusive, opaque, over-personalized, or misaligned if it is poorly timed or insufficiently grounded in the user's intent. Prior work on proactive conversational agents has similarly argued that the field must move beyond system capability alone and examine human-centered dimensions such as adaptivity, expectations, civility, and social implications \cite{deng2024humanproactive}. These concerns also resonate with foundational work on mixed-initiative interaction, trust in automation, and value-sensitive design, which emphasize that intelligent systems must support appropriate reliance, meaningful control, and explicit consideration of stakeholder values \cite{horvitz1999mixed,lee2004trust,hendry2021vsd}.

The First Workshop on Human-Centered Proactive and Personalized Agents for Interactive Information Access was organized around this emerging research problem. The workshop brought together perspectives from information retrieval, recommender systems~, human-computer interaction, dialogue systems, AI ethics, cognitive science, learning systems, and human-centered AI. Accepted contributions addressed knowledge-gap navigation, long-term conversational memory, proactive adaptive learning, AI-mediated care, proactive dialogue and co-creation, tiered transparency, value-sensitive design, implicit personalization, user welfare, and proactive recommendation. Across these contributions, the workshop treated proactivity not as a uniform increase in automation, but as a question of calibrated initiative: when should an agent act, what evidence should justify that action, how should the action be communicated, and how should users remain able to correct, contest, or refuse it?

The discussions further highlighted that evaluation remains a central open challenge~\cite{10.1145/3767695.3769484}. Traditional metrics such as relevance, accuracy, engagement, and task completion capture only part of what matters for proactive agents. Systems that initiate actions, surface unrequested information, or adapt over time must also be evaluated in terms of timing, initiative appropriateness, transparency, contestability, cognitive burden, trust calibration, privacy, and longitudinal impact. These dimensions are particularly important in interactive information access, where systems increasingly mediate not only access to information, but also the processes through which users learn, deliberate, make decisions, and revise their own understanding.

This report summarizes the workshop, its accepted contributions, and the main themes that emerged from the presentations and discussions. Rather than presenting a settled consensus, we use the report to document a research area in formation. We argue that human-centered proactive information access should be studied as both a technical capability and a socio-technical design problem: one concerned with what agents infer, when they intervene, how they explain their initiative, and how users retain agency in interactions that increasingly extend beyond the explicit query.

\section{Why Proactive Information Access, Why Now?}

The motivation for this workshop comes from a shift already visible across search, recommendation, and conversational systems: information access is becoming less episodic and more continuous. Users increasingly interact with systems that can preserve context, infer preferences, call tools, retrieve external information, and adapt across turns or sessions. This creates new possibilities for support that is not limited to answering explicit queries, but can also clarify needs, surface missing considerations, recommend next steps, and assist with longer-term goals.

This shift also makes proactivity a central design problem. Prior work in information retrieval has long recognized that users often begin with incomplete or evolving information needs \cite{belkin1980ask,bates1989berrypicking}, and conversational information seeking further frames access as an interactive process of refinement and sensemaking \cite{zamani2023conversational}. Agentic systems make this insight operational: if the system can observe context, maintain memory, and act across interactional turns, it may also decide when to take initiative. However, initiative is not automatically beneficial. A timely suggestion may reduce cognitive burden or reveal a knowledge gap, while a poorly grounded intervention may interrupt, over-personalize, or steer the user away from their own goals.

The workshop therefore approached proactive information access as a question of calibrated initiative. Rather than asking only how to make agents more capable or more autonomous, the workshop asked when initiative is warranted, what evidence should justify it, how it should be communicated, and how users should remain able to contest or redirect it. This framing connects recent work on human-centered proactive conversational agents \cite{deng2024humanproactive} with broader concerns around mixed-initiative interaction, trust, transparency, and value-sensitive design \cite{horvitz1999mixed,lee2004trust,hendry2021vsd}.

\section{Workshop Overview}

The First Workshop on Human-Centered Proactive and Personalized Agents for Interactive Information Access was held in conjunction with CHIIR 2026. The workshop was designed as an interdisciplinary forum for researchers studying proactive and personalized forms of information access across search, recommendation, conversational systems, learning support, and agentic AI. Its central goal was to examine how systems that remember, infer, recommend, clarify, or intervene can be designed in ways that remain human-centered, context-sensitive, and accountable to users.

The workshop combined invited talks, accepted paper presentations, and open discussion. This format was chosen to support both research presentation and community synthesis. The accepted papers reflected a broad range of perspectives, including knowledge-gap navigation, long-term conversational memory, proactive adaptive learning, AI-mediated care, proactive dialogue and co-creation, tiered transparency, value-sensitive design, implicit personalization, user welfare, and proactive recommendation. Together, these contributions showed that proactive information access is emerging as a shared concern across multiple research communities rather than a narrow technical subproblem.

A defining feature of the workshop was its emphasis on discussion. Rather than treating proactivity as a fixed system capability, participants examined it as a situated interactional decision: when should an agent take initiative, what should it know or infer before doing so, how should it explain its intervention, and how should users retain control? The discussion brought together technical questions around memory, user modeling, and evaluation with socio-technical questions around autonomy, privacy, trust, care, dependence, and human flourishing.

\section{Accepted Contributions}

The workshop received contributions that reflected the breadth of emerging work on proactive and personalized agents. Several papers examined proactivity as a question of initiative and intervention. These included work on human-centered proactivity in information access, proactive adaptive learning systems, knowledge-gap navigation under unknown unknowns, proactive dialogue systems for human-AI co-creation, and value-sensitive design for initiative decisions. Across these papers, proactivity was treated not simply as acting before the user asks, but as a design problem involving timing, uncertainty, user intent, reversibility, and control.

A second set of contributions focused on the infrastructure and risks of longitudinal personalization. Work on long-term conversational memory proposed mechanisms for maintaining temporally consistent user context across sessions, while papers on covert personalization and metadata-based engagement analysis examined how user models can emerge over time through sustained interaction. These contributions foregrounded both the promise and risk of personalization: memory and behavioral signals can enable more relevant support, but they can also introduce privacy concerns, misperception, dependence, and opaque adaptation.

A third set of papers addressed the human and social consequences of proactive agents. Contributions on AI-mediated care, tiered transparency, user welfare, and proactive recommendation highlighted how agents increasingly participate in emotionally, cognitively, and socially meaningful information practices. These papers emphasized that proactive systems must be evaluated not only by task performance or engagement, but also by how they affect user agency, trust, wellbeing, and the ability to understand or contest system behavior.

Together, the accepted contributions positioned proactive information access as a cross-cutting research area spanning search, recommendation, dialogue, learning, care, transparency, memory, and human-centered AI. They also provided the basis for the workshop's broader discussion: if agents are to move beyond reactive information delivery, the field must develop shared language, design principles, and evaluation methods for deciding when proactive support is appropriate and how it should be bounded.

\section{Emerging Themes from the Workshop}

\subsection{Proactivity as Calibrated Initiative}

A recurring theme across the workshop was that proactivity should not be equated with acting earlier, acting more often, or maximizing automation. Instead, participants converged on the idea of proactivity as calibrated initiative: the system must decide whether an intervention is warranted, what form it should take, and how much agency should remain with the user. This framing appeared across papers on knowledge-gap navigation, adaptive learning, proactive dialogue, and value-sensitive design, where initiative was tied to timing, uncertainty, reversibility, and user intent rather than prediction alone. It also shaped the discussion around interruptions: a proactive agent may support users by surfacing missing considerations or redirecting attention, but the same behavior can become intrusive if it is poorly timed or insufficiently grounded.

\subsection{Memory, Personalization, and Implicit Inference}

Long-term personalization emerged as both an enabling condition and a source of risk. Several contributions examined how proactive agents can use memory, behavioral traces, and inferred user attributes to provide more coherent support over time. At the same time, participants raised concerns about what agents should remember, what they may infer, and how those inferences should be updated, forgotten, or exposed to users. This tension was especially visible in discussions of long-term conversational memory, covert personalization, metadata-based engagement analysis, and user modeling. The workshop therefore treated memory not merely as technical infrastructure, but as a site where privacy, temporal consistency, user control, and personalization intersect.

\subsection{Values, Transparency, and Contestability}

Another major theme was the need to make the values embedded in proactive systems more explicit. Papers and discussions on value-sensitive design, tiered transparency, trust, and AI-mediated care emphasized that proactive agents do not only provide information; they make decisions about when to interrupt, what to prioritize, what assumptions to use, and how much to disclose. Participants repeatedly returned to questions such as: why now, why this suggestion, based on what data, and with what alternatives? These questions point to the importance of contestability. Human-centered proactivity requires mechanisms through which users can understand, correct, suppress, or redirect proactive behavior, rather than being passively acted upon by the system.

\subsection{User Welfare and Longitudinal Effects}

The workshop also highlighted that proactive agents may affect users over extended periods of interaction. Contributions on AI-mediated care, deep engagement behavior, adaptive learning, and implicit personalization raised questions about dependence, overuse, emotional reliance, cognitive burden, and wellbeing. The discussion made clear that engagement alone is an insufficient success criterion. A system that increases interaction time may not necessarily support user welfare, learning, or agency. For proactive agents, the relevant unit of analysis may need to shift from a single helpful response to the cumulative effects of repeated interventions over time.

\subsection{Evaluation Beyond Accuracy}

Finally, participants emphasized that existing evaluation practices are not sufficient for proactive and personalized agents. Accuracy, relevance, and task completion remain important, but they do not capture whether an agent intervened at the right time, respected user intent, explained its reasoning, avoided overreach, or supported long-term goals. The workshop surfaced several candidate evaluation dimensions, including initiative appropriateness, timing, transparency, contestability, privacy boundaries, trust calibration, cognitive burden, and longitudinal impact. Developing such evaluation frameworks is essential if proactive agents are to be assessed as interactive systems rather than as isolated response generators.

\section{Research Agenda}

The workshop discussions suggest several directions for future research on human-centered proactive information access.

\textbf{1. Model initiative as a first-class design decision.}
Future systems should make explicit decisions about when to remain passive, when to clarify, when to recommend, when to warn, and when to act. Treating initiative as a design variable requires attention to timing, confidence, stakes, reversibility, and user intent. This also calls for system architectures that can distinguish between low-risk suggestions and higher-commitment interventions.

\textbf{2. Develop evaluation methods for initiative appropriateness.}
Proactive agents cannot be evaluated only by the quality of their final output. A response may be correct but mistimed, useful but intrusive, or personalized but misaligned. Future benchmarks and user studies should measure whether interventions are warranted, whether they occur at appropriate moments, and whether users perceive them as helpful, controllable, and aligned with their goals.

\textbf{3. Design memory with boundaries.}
Long-term memory is central to personalization, but memory also introduces risks around privacy, persistence, misinference, and dependence. Future work should study mechanisms for temporal updating, forgetting, provenance tracking, user inspection, and selective suppression of remembered information. Memory should not be treated only as a way to improve relevance, but as an accountable component of interaction design.

\textbf{4. Support knowledge-gap navigation without derailing intent.}
A major opportunity for proactive agents is to help users identify missing constraints, alternative framings, or unknown unknowns. However, surfacing additional information can also distract from the user’s immediate task. Future systems should investigate how to expand the user’s information space in bounded ways, such as through clarifying questions, optional pivots, evidence bundles, or reversible suggestions.

\textbf{5. Make values, explanations, and contestability operational.}
Human-centered proactivity requires more than high-level commitments to transparency or user control. Future work should develop concrete interaction mechanisms such as ``why now'' explanations, interruption budgets, permission ladders, adjustable proactiveness levels, and user-facing controls for correction, refusal, and feedback. These mechanisms can help users understand not only what an agent recommends, but why it intervened at a particular moment.

\textbf{6. Study longitudinal effects on users.}
The impact of proactive agents may only become visible over time. Future research should examine how repeated interventions affect trust, agency, learning, reliance, cognitive burden, emotional attachment, and wellbeing. This requires moving beyond single-session studies toward longitudinal, ecologically valid evaluations that capture how users adapt to proactive systems in everyday information practices.

Together, these directions point toward a broader research agenda: proactive information access should be studied as an interactional and socio-technical problem, not only as a problem of prediction or automation. The central challenge is to design agents that can expand users' access to relevant information while preserving their ability to interpret, contest, and direct the interaction.

\section{Conclusion}

The First Workshop on Human-Centered Proactive and Personalized Agents for Interactive Information Access brought together researchers studying how agentic systems can move beyond reactive information delivery toward more adaptive, contextual, and initiative-taking forms of support. Across invited talks, paper presentations, and open discussion, the workshop highlighted that proactivity is not a single capability, but a set of design decisions about timing, grounding, personalization, transparency, contestability, and user welfare.

A central outcome of the workshop was the recognition that proactive information access requires a broader evaluative vocabulary. Systems that remember, infer, recommend, or intervene must be assessed not only by relevance, accuracy, or engagement, but also by whether their initiative is appropriate, understandable, bounded, and aligned with user goals. This is especially important as proactive agents begin to operate across longer interactions, richer user models, and more consequential domains.

The workshop marks an early step toward building a community around human-centered proactive information access. Future work will require closer connections between information retrieval, recommender systems, HCI, dialogue systems, cognitive science, AI ethics, and human-centered AI. By documenting the themes and challenges surfaced through the workshop, this report aims to support a shared research agenda for proactive agents that expand access to information while preserving human agency, privacy, and meaningful control.
\bibliography{sigirforum}

\end{document}